\documentstyle[12pt,a4]{article}
\begin{document}
 \setlength{\baselineskip}{8.5mm}

\hfill {\large NU961}
\vskip 2.0cm
\begin{center}
 {\LARGE Comments on "Analysis of Two-Body Decays of
  Charmed Baryons \\
    Using the Quark-Diagram Scheme"}
\end{center}
\vspace{1.5cm}
\begin{center}
 {\large Yoji Kohara} \\
  Nihon University at Shonan, Fujisawa, Kanagawa 252, Japan
\end{center}
\vspace{1.5cm}
\begin{center}
 Abstract
\end{center}

 It is proved that the nonleptonic decay amplitudes of antitriplet
 charmed baryons to octet baryons do not depend on the
 representation of the octet baryons in the quark diagram scheme.
 Relations among various representations are derived, and correct
 decay amplitudes of the charmed baryons are given.

\vspace{1cm} 
\begin{flushleft}
PACS numbers: 12.15.Ji, 12.39.-x, 13.30.-a, 13.30.Eg, 
\end{flushleft}

\newpage

  The quark diagram scheme was applied to the two-body
  nonleptonic decays of charmed  baryons in Ref. \cite{ko}
  and Ref. \cite{ch}.  The wave
  functions of SU(3) octet baryons can be represented in some
  different  forms. However, the  physics is independent of
  the convention one chooses. In this paper it  will be
  proved that both schemes are equivalent.
  
 We consider only the decays of antitriplet charmed baryons
 to an octet baryon and a nonet pseudoscalar meson here.
 The octet baryon states can be represented in terms of the
 quark states that are antisymmetric in the first and the
 second quarks.    
\begin{eqnarray}
 |\psi^k(8)_{A12}\rangle &=&  \eta_k~  |[q_aq_b]q_a
  \rangle ~~~~~ 
  \mbox{for}~ k=p,n,\Sigma^+,\Sigma^-,\Xi^0,\Xi^-,\\
 |\psi^{\Sigma^0}(8)_{A12}\rangle & =& \frac{1}{\sqrt{2}}
   |[sd]u\rangle + \frac{1}{\sqrt{2}} |[su]d\rangle,\\
 |\psi^\Lambda(8)_{A12}\rangle
 & =&  - \frac{1}{\sqrt{6}} |[sd]u\rangle +
  \frac{1}{\sqrt{6}} |[su]d\rangle + \frac{2}{\sqrt{6}}
 |[du]s\rangle ,
\end{eqnarray}
where 
\begin{equation}
 |[q_aq_b]q_c\rangle = \frac{1}{\sqrt{2}} 
  (|q_aq_bq_c\rangle - |q_bq_aq_c\rangle),
\end{equation}
and $\eta_k$ is a phase factor which depends on the convention.
The octet baryon states can be represented also by
antisymmetrizing the second and the third quarks.
\begin{eqnarray}
 |\psi^k(8)_{A23}\rangle &
  =& \eta_k~  |q_a[q_aq_b]\rangle~~~~~~~  
  \mbox{for}~ k=p,n,\Sigma^+,\Sigma^-,
  \Xi^0,\Xi^-,\\
 |\psi^{\Sigma^0}(8)_{A23}\rangle
 & =& \frac{1}{\sqrt{2}} |u[sd]\rangle +
  \frac{1}{\sqrt{2}} |d[su]\rangle,\\ 
 |\psi^\Lambda(8)_{A23}\rangle
  & =& - \frac{1}{\sqrt{6}} |u[sd]\rangle +
  \frac{1}{\sqrt{6}} |d[su]\rangle + \frac{2}{\sqrt{6}}
 |s[du]\rangle ,
\end{eqnarray}
where
\begin{equation}
 |q_a[q_bq_c]\rangle = \frac{1}{\sqrt{2}}(|q_aq_bq_c\rangle
  -|q_aq_cq_b\rangle).
\end{equation}
Similarly, there are also the representations
$|\psi^k(8)_{A31}\rangle$ that are antisymmetric in the first
and the third quarks. As we are considering only octet states,
the following relation holds among quark antisymmetric states:
\begin{equation}
 |[q_aq_b]q_c\rangle + |q_c[q_aq_b]\rangle
 + |[q_bq_cq_a]\rangle = 0,
\end{equation}
where
\begin{equation}
 |[q_bq_cq_a]\rangle = \frac{1}{\sqrt{2}} (|q_bq_cq_a\rangle
 - |q_aq_cq_b\rangle ).
\end{equation}
This relation results from that the left hand side is totally
antisymmetric. Because of this equation two of three terms are
independent. We take the first and the second terms as
independent bases. Other relations are also derived from the
condition of octet states.
\begin{equation}
 |[q_aq_b]q_c\rangle + |[q_bq_c]q_a\rangle
 + |[q_cq_a]q_b\rangle = 0,
\end{equation}
\begin{equation}
 |q_a[q_bq_c]\rangle + |q_b[q_cq_a]\rangle
 + |q_c[q_aq_b]\rangle = 0.
\end{equation}
Then the states $|[q_aq_b]q_c\rangle$ and  $|q_c[q_aq_b]\rangle$
constitute a complete set for octet state space under the 
above eqs. (11) and (12), though they are not orthogonal. 
However, it is not essential in our issue whether the bases
are orthogonal or not. The antisymmetric octet baryon states 
$|\psi^k(8)_{A12}\rangle $and $|\psi^k(8)_{A23}\rangle$
can be represented in terms of quark antisymmetric states 
$|[q_aq_b]q_c\rangle$ and $|q_a[q_bq_c]\rangle$, respectively.
\begin{eqnarray}
 |\psi^k(8)_{A12}\rangle &=& \sum_{q_a,q_b,q_c}
 |[q_aq_b] q_c\rangle~
 \langle [q_aq_b]q_c|\psi^k(8)_{A12}\rangle~,   \\
 |\psi^k(8)_{A23}\rangle &=& \sum_{q_a,q_b,q_c}
 |q_c[q_aq_b]\rangle~
 \langle q_c[q_aq_b]|\psi^k(8)_{A23}\rangle~.
\end{eqnarray}
These equations do not depend on whether the bases are
orthogonal or not. Most generally, the octet baryon states
$|B^k(8)\rangle$ can be written
as a combination of $|\psi^k(8)_{A12}\rangle$ and
$|\psi^k(8)_{A23}\rangle$
\begin{equation}
 |B^k(8)\rangle = \alpha~ |\psi^k(8)_{A12}\rangle
 + \beta~ |\psi^k(8)_{A23}\rangle.
\end{equation}
We suppressed spin wave functions for simplicity. This
equation corresponds to the following equation of Ref. [2]
\begin{equation}
 |B^k(8)\rangle = a~ |\psi^k(8)_{A}\rangle
 + b~ |\psi^k(8)_{S}\rangle,
\end{equation}
where $|\psi^k(8)_S\rangle$ and $|\psi^k(8)_A\rangle$
denote the octet baryon states that are symmetric and
antisymmetric in the first two quarks, respectively.
The presice values of $\alpha$ and $\beta$
are not important in our issue.

The decay amplitudes of the antitriplet charmed baryon 
$B_c^{i_0}$ to an octet baryon $B^{k_0}(8)$ and an octet
pseudoscalar meson $M^{j_0}(8)$ are given as follows:
\begin{eqnarray}
 A(i_0\rightarrow j_0~k_0) &=& \langle B^{i_0}_c|\hat 
 H_W|M^{j_0}(8)
 \rangle~|B^{k_0}(8)\rangle   \nonumber \\ 
 ~~~~~~&=& \langle B^{i_0}_c|\hat H_W|M^{j_0}(8)\rangle~
 \Big(\alpha~ |\psi^{k_0}(8)_{A12}\rangle +
 \beta~|\psi^{k_0}(8)_{A23}\rangle\Big) \nonumber   \\
 &=& \sum_{q_i} \alpha~ \langle B^{i_0}_c|\hat H_W|M^{j_0}(8)
 \rangle~|[q_1q_2]q_3\rangle~\langle [q_1q_2]q_3
 |\psi^{k_0}(8)_{A12}\rangle   \nonumber \\
 &+& \sum_{q_i} \beta~ \langle B^{i_0}_c|\hat H_W
 |M^{j_0}(8)\rangle~
 |q_3[q_1q_2]\rangle~\langle q_3[q_1q_2]|\psi^{k_0}(8)_{A23}
 \rangle~~ \nonumber \\ 
 &=& \!\!\!\sum_{\bar q,q',q_i}~\!\! \alpha
 \,\langle B^{i_0}_c 
 |\hat H_W|\bar q q'\rangle~|[q_1q_2]q_3\rangle~  
 \langle\bar {q}q'|\phi^{j_0}(8)\rangle~
 \langle [q_1q_2]q_3|\psi^{k_0}(8)_{A12}\rangle  
 \nonumber \\
 &+& \!\!\!\sum_{\bar q,q',q_i} \beta\,\langle B^{i_0}_c 
 |\hat H_W|\bar{q} q'\rangle~|q_3[q_1q_2]\rangle~  
 \langle \bar{q}q'|\phi^{j_0}(8)\rangle~
 \langle q_3[q_1q_2]|\psi^{k_0}(8)_{A23}\rangle. 
\end{eqnarray}
In this case there are nine kinds of the quark diagrams as
depicted in Fig. 1. Type $D$, $D'$ and $G$ consist of two
diagrams, one diagram in which the first and the second quarks
are antisymmetrized and one diagram in which the
second and the third quarks are antisymmetrized.
On the other hand type $A, B, C, E, F$ and $H$ consist of 
only one diagram, because the spectator quarks stay
antisymmetric in the final baryons or the quarks produced in
the weak interaction are antisymmetric due to the Pati-Woo
theorem \cite{ko,k}. Then we reach the following expression
of the decay amplitudes:
\begin{eqnarray}
 A(i_0\rightarrow j_0~k_0) &=&\!\!\!\sum_{\bar q,q',q_i}~\!\!
 \bigl\{  ~A~\langle\bar{q}_0 q'_0|\phi^{j_0}(8)\rangle~\langle 
 [q_1q_2]q_3|\psi^{k_0}(8)_{A12}\rangle~ \nonumber \\
  &+& B~\langle\bar q_0 q_3|\phi^{j_0}(8)\rangle~\langle 
 [q_1q_2]q'_0|\psi^{k_0}(8)_{A12}\rangle~ \nonumber \\
 &+& C~\langle\bar q_0 q_2|\phi^{j_0}(8)\rangle~\langle 
 [q'_0q_3]q_1|\psi^{k_0}(8)_{A12}\rangle~ \nonumber \\
  &+& D_{1}~\langle\bar q_0 q_3|\phi^{j_0}(8)\rangle~\langle 
 [q_1q'_2]q_0|\psi^{k_0}(8)_{A12}\rangle~ \nonumber \\
 &+& D_{2}~\langle\bar q_0 q_3|\phi^{j_0}(8)\rangle~\langle 
 q_1[q'_2q_0]|\psi^{k_0}(8)_{A23}\rangle \nonumber \\
 &+& D'_{1}~\langle\bar q_0 q'_2|\phi^{j_0}(8)\rangle~\langle 
 [q_1q_3]q_0|\psi^{k_0}(8)_{A12}\rangle \nonumber \\
  &+& D'_2~\langle\bar q_0q'_2|\phi^{j_0}(8)\rangle~\langle 
 q_1[q_3q_0]|\psi^{k_0}(8)_{A23}\rangle \nonumber \\
 &+& E~\langle\bar q_0 q_2|\phi^{j_0}(8)\rangle~\langle 
 [q'_1q_3]q_0|\psi^{k_0}(8)_{A12}\rangle \nonumber \\
  &+& F~\langle\bar q_0 q_3|\phi^{j_0}(8)\rangle~\langle 
 [q_1q_2]q_0|\psi^{k_0}(8)_{A12}\rangle \nonumber \\
 &+& G_1~\langle\bar q_0 q_2|\phi^{j_0}(8)\rangle~\langle 
 [q_3q_1]q_0|\psi^{k_0}(8)_{A12}\rangle \nonumber \\
  &+& G_2~\langle\bar q_0 q_2|\phi^{j_0}(8)\rangle~\langle 
 q_3[q_1q_0]|\psi^{k_0}(8)_{A23}\rangle \nonumber \\  
 &+& H~\langle\bar q_0 q_0|\phi^{j_0}(8)\rangle~\langle
 [q'_2q_3]q_1|\psi^{k_0}(8)_{A12}\rangle\bigr\}.
\end{eqnarray}
Each term corresponds to each diagram of Fig. 1.
From the above equation the decay amplitudes are represented 
in terms of twelve parameters ($A$ etc.). These parameters
are essentially the same with the parameters $a$ etc. in
Ref. [1], but they differ by only numerical factors. The
relations between two sets of parameters are as follows:
\begin{eqnarray}
 A=\frac{8}{\sqrt{6}}~a, & \qquad
 \displaystyle{B=\frac{8}{\sqrt{6}}~b,} &
 \qquad  C=\frac{4}{\sqrt{6}}~c, \nonumber \\
 D_1=\frac{4}{\sqrt{6}}~d_1, & \qquad
 \displaystyle{D_2= - \frac{4}{\sqrt{6}}~d_2,} &
 \qquad  D'_1=\frac{4}{\sqrt{6}}~d_3, \nonumber \\ 
 D'_2= - \frac{4}{\sqrt{6}}~d_4, & \qquad
 \displaystyle{E=\frac{4}{\sqrt{6}}~e,}&
 \qquad  H= - \frac{4}{\sqrt{6}}~h.  
\end{eqnarray}
With these new parameters we can more easily compare 
with the amplitudes in Ref. [2].

The relations between $|\psi^k(8)_{A12}\rangle,
|\psi^k(8)_{A23}\rangle$ in this paper and
$|\psi^k(8)_A\rangle, |\psi^k(8)_S\rangle$
in Ref. [2] are as follows:
\begin {eqnarray}
|\psi^k(8)_A\rangle& =& |\psi^k(8)_{A12}\rangle, \\
|\psi^k(8)_S\rangle &=& -\frac{1}{\sqrt{3}} 
|\psi^k(8)_{A12}\rangle
 -\frac{2}{\sqrt{3}} |\psi^k(8)_{A23}\rangle,
\end{eqnarray}
for all $k$. From Eqs. (20) and (21) octet baryon states
are expressed as
\begin{eqnarray}
 |B^k(8)\rangle & = & a~|\psi^k(8)_A\rangle 
 + b~|\psi^k(8)_S\rangle  \nonumber \\
   & = & (a-\frac{1}{\sqrt{3}} b ) |\psi^k(8)_{A12}\rangle
          -\frac{2}{\sqrt{3}} b~|\psi^k(8)_{A23}\rangle.
\end{eqnarray}
Comparing with Eq. (15), we obtain
\begin{equation}
 \alpha = a- \frac{1}{\sqrt{3}}~b, \;\;\;\;\;\; 
 \beta = -\frac{2}{\sqrt{3}}~b.
\end{equation}
Therefore the relations between the above parameters
$A$ etc. and the parameters ${\cal A}_A$ etc. in Ref. [2] 
are written as follows:
\begin{eqnarray}
  A={\cal A}_A, & \qquad B={\cal B}'_A, & \qquad
  C={\cal B}_A, \nonumber \\
  D_1={\cal C}'_A-\frac{1}{\sqrt{3}}~{\cal C}'_S, & \qquad
  \displaystyle{D_2=-\frac{2}{\sqrt{3}}~{\cal C}'_S,} &
  \qquad  D'_1={\cal C}_{2A}-\frac{1}{\sqrt{3}}~
  {\cal C}_{2S}, \nonumber \\
  D'_2=-\frac{2}{\sqrt{3}}~{\cal C}_{2S}, & \qquad
  E={\cal C}_{1A}, & \qquad
  F={\cal E}'_A, \nonumber \\
  G_1={\cal E}_A-\frac{1}{\sqrt{3}}~{\cal E}_S,& \qquad
  \displaystyle{G_2=-\frac{2}{\sqrt{3}}{\cal E}_S.} &
\end{eqnarray}
These relations are the same with Eq. (63) in Ref. [2], but 
they were pointed out first in Ref. \cite{k2}.
The quark diagram schemes in Ref. [1] and in Ref. [2] are
equivalent under these relations.
There are many errors in the decay amplitudes of Ref. [2].
The correct decay amplitudes
without final state interactions are given in terms of the
parameters of this paper in Tables I, II and III.
They can be translated into Chau's amplitudes using Eqs. (24).

We can also use symmetric bases. We define
\begin{equation}
|\{q_aq_b\}q_c\rangle = \frac{1}{\sqrt{2}(1-\delta_{ab})+
 2\delta_{ab}} (|q_aq_bq_c\rangle + |q_bq_aq_c\rangle),
\end{equation}
\begin{equation}
|q_c\{q_aq_b\}\rangle = \frac{1}{\sqrt{2}(1-\delta_{ab})+
2\delta_{ab}} (|q_cq_aq_b\rangle + |q_cq_bq_a\rangle),
\end{equation}
\begin{equation}
 |\{q_aq_cq_b\}\rangle  = \frac{1}{\sqrt{2}(1-\delta_{ab})+
 2\delta_{ab}} (|q_aq_cq_b\rangle + |q_bq_cq_a\rangle).
\end{equation}
The octet baryon states in which the first and the second quarks
are symmetric are written as
\begin{eqnarray}
 |\psi^k(8)_{S12}\rangle
  &=& \frac{1}{\sqrt{3}} \eta_k |\{q_aq_b\}q_a\rangle
  -\frac{2}{\sqrt{6}} \eta_k ~|q_aq_aq_b\rangle 
 ~~~~\mbox{for}~k= p, n, \Sigma^+,\Sigma^-,\Xi^0,\Xi^-, \\
 |\psi^{\Sigma^0}(8)_{S12}\rangle
  &=&- \frac{1}{\sqrt{6}} |\{sd\}u\rangle -\frac{1}{\sqrt{6}}
 |\{su\}d\rangle +\frac{2}{\sqrt{6}}|\{du\}s\rangle, \\
 |\psi^\Lambda(8)_{S12}\rangle
 &=& -\frac{1}{\sqrt{2}} |\{sd\}u\rangle + \frac{1}{\sqrt{2}}
 |\{su\}d\rangle.
\end{eqnarray}
The symmetric octet states $|\psi^k(8)_{S23}\rangle$ and
$|\psi^k(8)_{S31}\rangle$ are defined in the same way.
As we are treating only SU(3) octet states,
the following equation holds.
\begin{equation}
    |\{q_aq_b\}q_c\rangle + |q_c\{q_aq_b\}\rangle
    + |\{q_aq_cq_b\}\rangle = 0.
\end{equation}
So two of three terms are independent. We adopt  
$|\{q_aq_b\}q_c\rangle$ and $|q_c\{q_aq_b\}\rangle$
as independent bases. Under the octet conditions
\begin{equation}
 |\{q_aq_b\}q_c\rangle + |\{q_bq_c\}q_a\rangle
 + |\{q_cq_a\}q_b\rangle = 0
\end{equation}
and
\begin{equation}
 |q_a\{q_bq_c\}\rangle + |q_b\{q_cq_a\}\rangle
 + |q_c\{q_aq_b\}\rangle = 0,
\end{equation}
$|\{q_aq_b\}q_c\rangle$ and $|q_a\{q_bq_c\}\rangle$
constitute a complete set. Though these are not orthogonal,
symmetric octet baryon states
$|\psi^k(8)_{S12}\rangle$ and $|\psi^k(8)_{S23}\rangle$ can be
written using quark symmetric bases $|\{q_aq_b\}q_c\rangle$
and $|q_c\{q_aq_b\}\rangle$ respectively.
\begin{eqnarray}
 |\psi^k(8)_{S12}\rangle &=& \sum_{q_a,q_b,q_c}|\{q_aq_b\} q_c
 \rangle~ \langle \{q_aq_b\}q_c|\psi^k(8)_{S12}\rangle~,   \\
 |\psi^k(8)_{S23}\rangle &=& \sum_{q_a,q_b,q_c}|q_c\{q_aq_b\}
 \rangle~ \langle q_c\{q_aq_b\}|\psi^k(8)_{S23}\rangle~.
\end{eqnarray}
Most general form of octet baryon states $|B^k(8)\rangle$ is a
 combination of the two parts.
\begin{equation}
 |B^k(8)\rangle = \alpha' ~|\psi^k(8)_{S12}\rangle
 + \beta' ~|\psi^k(8)_{S23}\rangle.
\end{equation}

The decay amplitudes of antitriplet charmed baryon $B_c^{i_0}$ 
to an octet baryon $B^{k_0}(8)$ and an octet pseudoscalar meson
$M^{j_0}(8)$ are written as follows:
\begin{eqnarray}
 A(i_0\rightarrow j_0 k_0) &=& 
  \!\!\!\sum_{\bar q,q',q_i}~\!\! \alpha'~\langle B^{i_0}_c 
  |\hat H_W|\bar q q'\rangle~|\{q_1q_2\}q_3\rangle~ 
  \langle\bar {q}q'|\phi^{j_0}(8)\rangle~
  \langle \{q_1q_2\}q_3|\psi^{k_0}(8)_{S12}\rangle \nonumber \\
  &+& \!\!\!\sum_{\bar q,q',q_i} \beta'~\langle B^{i_0}_c 
  |\hat H_W|\bar{q} q'\rangle~|q_3\{q_1q_2\}\rangle~ 
  \langle \bar{q}q'|\phi^{j_0}(8)\rangle~
  \langle q_3\{q_1q_2\}|\psi^{k_0}(8)_{S23}\rangle. 
\end{eqnarray}
In this case there are nine kinds of quark diagrams as
 depicted in Fig. 2. Each type consists of two diagrams,
 one diagram in which the first and the second quarks are
 symmetrized and one diagram in which the second and the third
 quarks are symmetrized. In type $A_S, B_S$ and $F_S$ diagrams
 the spectator quarks $q_1$ and $q_2$ are antisymmetric and in
 type $C_S$, $E_S$ and $H_S$ diagrams the quarks produced in
 the weak interaction are antisymmetric, so we must take the
 difference of two diagram contributions. Then the expression
 of the decay amplitudes is
\begin{eqnarray}
 A(i_0\rightarrow j_0 k_0) &=&\!\!\!\sum_{\bar q,q',q_i}~\!\!
 \Bigl\{A_S~\langle\bar{q}_0 q'_0|\phi^{j_0}(8)\rangle \bigl( 
 \langle \{q_2q_3\} q_1|\psi^{k_0}(8)_{S12}\rangle~ - \langle
 q_2\{q_3q_1\}|\psi^{k_0}(8)_{S23}\rangle \bigr) \nonumber \\
  &+& B_S~\langle\bar q_0 q_3|\phi^{j_0}(8)\rangle \bigl( 
 \langle \{q_2q'_0\}q_1|\psi^{k_0}(8)_{S12}\rangle~ 
  -  \langle q_2\{q'_0q_1\}|\psi^{k_0}(8)_{S23}\rangle \bigr) 
 \nonumber \\
  &+& C_S~\langle\bar q_0 q_2|\phi^{j_0}(8)\rangle \bigl( 
 \langle \{q_3q_1\}q'_0|\psi^{k_0}(8)_{S12}\rangle~
   -  \langle q_3\{q_1q'_0\}|\psi^{k_0}(8)_{S23}\rangle \bigr)
 \nonumber \\
  &+& D_{S1}~\langle \bar q_0 q_3|\phi^{j_0}(8)\rangle\langle 
  \{q'_2q_0\}q_1|\psi^{k_0}(8)_{S12}\rangle~  \nonumber   \\
  &+& D_{S2}~\langle\bar q_0 q_3|\phi^{j_0}(8)\rangle\langle 
  q'_2\{q_0q_1\}|\psi^{k_0}(8)_{S23}\rangle   \nonumber \\
  &+& D'_{S1}~\langle\bar q_0 q'_2|\phi^{j_0}(8)\rangle\langle 
  \{q_3q_0\}q_1|\psi^{k_0}(8)_{S12}\rangle   \nonumber \\
  &+& D'_{S2}~\langle\bar q_0 q'_2|\phi^{j_0}(8)\rangle\langle 
  q_3\{q_0q_1\}|\psi^{k_0}(8)_{S23}\rangle   \nonumber \\
  &+& E_S~ \langle\bar q_0 q_2|\phi^{j_0}(8)\rangle \bigl(\langle 
  \{q_3q_0\}q'_1|\psi^{k_0}(8)_{S12}\rangle  
  - \langle q_3\{q_0q'_1\}|\psi^{k_0}(8)_{S23}\rangle \bigr)
  \nonumber \\ 
  &+& F_S~\langle\bar q_0 q_3|\phi^{j_0}(8)\rangle \bigl( \langle 
  \{q_2q_0\}q_1|\psi^{k_0}(8)_{S12}\rangle    
  - \langle q_2\{q_0q_1\}|\psi^{k_0}(8)_{S23}\rangle \bigr)
  \nonumber \\
  &+& G_{S1}~\langle\bar q_0 q_2|\phi^{j_0}(8)\rangle\langle 
  \{q_1q_0\}q_3|\psi^{k_0}(8)_{S12}\rangle \nonumber \\
  &+& G_{S2}~\langle\bar q_0 q_2|\phi^{j_0}(8)\rangle\langle 
  q_1\{q_0q_3\}|\psi^{k_0}(8)_{S23}\rangle \nonumber \\  
  &+& H_S~\langle\bar q_0 q_0|\phi^{j_0}(8)\rangle \bigl( \langle
  \{q_3q_1\}q'_2|\psi^{k_0}(8)_{S12}\rangle 
  - \langle q_3\{q_1q'_2\}|\psi^{k_0}(8)_{S23}\rangle \bigr) \Bigr\}.
\end{eqnarray}
The factor $\{\sqrt{2}(1-\delta_{q_iq_j})+2\delta_{q_iq_j}\}$
was dropped in each term for simplicity.

The decay amplitudes were represented in terms of the above twelve
parameters. From the relations
\begin {equation}
 |\psi^k(8)_S\rangle = |\psi^k(8)_{S12}\rangle 
\end{equation}
and
\begin{equation}
 |\psi^k(8)_A\rangle = \frac{1}{\sqrt{3}}~|\psi^k(8)_{S12}\rangle
          +\frac{2}{\sqrt{3}}~|\psi^k(8)_{S23}\rangle,
\end{equation}
octet baryon states $|B^k(8)\rangle$ are written as
\begin {eqnarray}
 |B^k(8)\rangle &=& a~ |\psi^k(8)_A\rangle + b~|\psi^k(8)_S\rangle
  \nonumber \\
   &=& (\frac{1}{\sqrt{3}}~a+ b )~|\psi^k(8)_{S12}\rangle
          +\frac{2}{\sqrt{3}}a~|\psi^k(8)_{S23}\rangle.
\end{eqnarray}
Relations between the above parameters ($A_S$ etc.) and
the parameters (${\cal A}_A$ etc.) in Ref. [2] are derived
from this equation.
\begin{eqnarray}
 A_S=\frac{1}{\sqrt{3}}{\cal A}_A, & \qquad
  \displaystyle{B_S=\frac{1}{\sqrt{3}}{\cal B}'_A,} & \qquad
 C_S=\frac{1}{\sqrt{3}}{\cal B}_A, \nonumber \\ 
  D_{S1}={\cal C}'_S+\frac{1}{\sqrt{3}}{\cal C}'_A, &
 \qquad \displaystyle{D_{S2}=\frac{2}{\sqrt{3}}{\cal C}'_{S},}
 & \quad D'_{S1}={\cal C}_{2A}+\frac{1}{\sqrt{3}}{\cal C}_{2S}, 
 \nonumber \\
  D'_{S2}=\frac{2}{\sqrt{3}}{\cal C}_{2S}, & \qquad
 \displaystyle{E_S=\frac{1}{\sqrt{3}}{\cal C}_{1A},} & \qquad
  G_{S1}={\cal E}_A+\frac{1}{\sqrt{3}}{\cal E}_S, \nonumber \\ 
 G_{S2}=\frac{2}{\sqrt{3}}{\cal E}_S, & \qquad 
 \displaystyle{F_S=\frac{1}{\sqrt{3}}{\cal E}'_A.} &
\end{eqnarray}  
Thus this symmetric scheme is also equivalent to Chau's scheme
under these relations.

In conclusion, though octet baryon states can be represented
in some different forms, they lead to equivalent decay amplitudes
in the quark diagram scheme. The relations among the different
representations were obtained like Eqs. (24) and (42).

\newpage

\def\lamc{\Lambda^+_c}
\def\xia{\Xi^{+}_c}
\def\xi0a{\Xi^{0}_c}
\def\sqa{{\frac{1}{\sqrt{2}}}}
\def\sqb{{\frac{1}{\sqrt{3}}}}
\def\sqc{{\frac{1}{\sqrt{6}}}}
\def\sqd{{\frac{1}{2\sqrt{3}}}}
\def\sqe{{\frac{1}{2\sqrt{2}}}}
\def\sqf{{\frac{1}{2\sqrt{6}}}}
\def\fa{{\frac{1}{2}}}
\def\fb{{\frac{1}{3}}}
\def\fc{{\frac{1}{6}}}
\begin{center}
 TABLE I. Cabibbo-allowed decay amplitudes
\end{center}
\[ \begin{array}{rllllll}
\hline \hline & & & & & & \\
\mbox{Decay} & \mbox{modes} & & &\qquad\qquad 
\mbox{Amplitudes}~~\quad & & \\
& & & & & &  \\
\hline & & & & & & \\
\lamc  \to & p \bar{K}^0  & & &\fa (B + D_2) &  & \\
& & & & & & \\
\to &\Lambda\pi^+  & & & - \sqf (2A + C+D'_1+D'_2-E)
 & & \\ & & & & & & \\
\to &\Sigma^0\pi^+ & & &\sqe(C+D'_1-D'_2-E) & & \\
& & & & & & \\
\to &\Sigma^+\pi^0  & & &\sqe(-C-D'_1+D'_2+E) & &  \\
& & & & & & \\
\to &\Sigma^+\eta_1 & & &\sqd(C-D_2-D'_1+D'_2+E-3H) & & \\
& & & & & & \\
\to &\Sigma^+\eta_8 & & &\sqf(C+2D_2-D'_1+D'_2+E) && \\
& & & & & & \\
\to &\Xi^0 K^+ & & &\fa(-D'_1+E) & & \\
& & & & & & \\
\xia \to &\Sigma^+ \bar{K}^0 & & &\fa(B-C) & & \\
& & & & & & \\
\to &\Xi^0 \pi^+ & & &\fa(A+C) & & \\
& & & & & & \\
 \xi0a \to & \Lambda \bar{K}^0 & & &\sqf(-B-C+D_1-2D_2+E)
& & \\ & & & & & & \\
\to &\Sigma^0 \bar{K}^0  & & &\sqe(-B+C-D_1-E) && \\
& & & & & & \\ 
\to &\Sigma^+ K^-  & & &\fa(D_1+E) &&\\
& & & & & & \\
\to &\Xi^- \pi^+  & & & - \fa(A+D'_2) & & \\
& & & & & & \\
\to &\Xi^0 \pi^0  & & &\sqe(-C+D'_2) & & \\
& & & & & & \\
\to &\Xi^0 \eta_1 & & &\sqd(C+D_1-D_2+D'_2+E-3H)& & \\
& & & & & & \\
\to &\Xi^0 \eta_8 & & &\sqf(C-2D_1+2D_2+D'_2-2E)&&\\
& & & & & & \\
\hline
\end{array}
\]
\pagebreak
\begin{center}
 TABLE II. Cabibbo-suppressed decay amplitudes
\end{center}
\[ \begin{array}{rllllll}
\hline
\hline
& & & & & &  \\
\mbox{Decay}& \mbox{modes} & & & \qquad \qquad
\mbox{Amplitudes} \quad & & \\
& & & & & &  \\
\hline
& & & & & & \\
\lamc \to &\Sigma^0 K^+  & &&\sqe(C-D'_2-G_2) &&\\
& & & & & & \\
\to &\Sigma^+ K^0 & & &\fa(C+D_2-G_2) &&\\
& & & & & & \\
\to &\Lambda K^+  & & &-\sqf(2A+C-2D'_1+D'_2+2E+2F-
2G_1+G_2) & & \\
& & & & & & \\
\to & n \pi^+  & & &\fa(-A-C-D'_1+E+F-G_1) & & \\
& & & & & & \\
\to & p \pi^0 & & &\sqe(B-C+D_2-D'_1+D'_2+E+F-G_1) & & \\
& & & & & & \\
\to & p \eta_1 & & &\sqd(C-D_2-D'_1+D'_2+E+F
-G_1+2G_2-3H) &&\\
& & & & & & \\
\to & p \eta_8 & & &{\frac{1}{2\sqrt{6}}}
(-3B+C-D_2-D'_1+D'_2+E+F-G_1+2G_2)& & \\
& & & & & & \\
\xia \to &\Sigma^0 \pi^+  & & &
\sqe(A-D'_1+D'_2+E-F+G_1-G_2) &&\\
& & & & & & \\
\to &\Sigma^+ \pi^0 & & &\sqe(B+D'_1-D'_2-E+F-G_1+G_2)
 &&\\ & & & & & & \\
\to &\Sigma^+ \eta_1 & & &\sqd
(-C+D_2+D'_1-D'_2-E+F-G_1+2G_2+3H)& & \\
& & & & & & \\
\to &\Sigma^+ \eta_8 & & &\sqf
(-3B+2C-2D_2+D'_1-D'_2-E+F-G_1-G_2)&&\\
& & & & & & \\
\to &\Lambda \pi^+  & & &\sqf
(-A-2C+D'_1+D'_2-E+F-G_1-G_2)& & \\
& & & & & & \\
\to &\Xi^0 K^+ & & &\fa(A+C+D'_1-E+F-G_1)& & \\
& & & & & & \\
\to &p \bar{K}^0  & & &-\fa(C+D_2+G_2)& & \\
& & & & & & \\
\hline
\end{array}
\]
{\hsize 22.6 cm
\centerline{(continued)}}
\pagebreak
\[ \begin{array}{rllllll}
\hline
\hline
& & & & & &  \\
\mbox{Decay}& \mbox{modes} & & & \qquad \qquad
\mbox{Amplitudes} \quad  & & \\
& & & & & &  \\
\hline
& & & & & &  \\
\xi0a \to &\Sigma^+ \pi^- & & &-\fa(D_1+E+G_1-G_2) & &\\
& & & & & & \\
\to &\Sigma^- \pi^+ & & &-\fa(-A-D'_2+F) & & \\
& & & & & &  \\
\to &\Sigma^0 \pi^0 & & &-{\frac{1}{4}}
(B+D_1-D'_2+E+F+G_1-G_2) & &\\
& & & & & &  \\
\to &\Sigma^0 \eta_1 & & &\sqf
(-B+C+D_1-D_2+D'_2+E-F+G_1-2G_2-3H)& & \\
& & & & & & \\
\to &\Sigma^0 \eta_8  & & &{\frac{1}{4\sqrt{3}}}
(3B-2C+D_1+2D_2+D'_2+E-F+G_1+G_2)& &\\
& & & & & &  \\
\to &\Lambda \pi^0 & & &{\frac{1}{4\sqrt{3}}}
(-B+2C+D_1-2D_2-3D'_2+E-F+G_1+G_2)& &\\
& & & & & & \\
\to &\Lambda \eta_1 & & &{\frac{1}{6\sqrt{2}}}
(-3C-3D_1+3D_2-3D'_2-3E-F+G_1-2G_2+9H)&& \\
& & & & & & \\
\to &\Lambda \eta_8  & & &{\frac{1}{12}}
(3B+3D_1-3D'_2+3E-F-5G_1+G_2)& & \\
& & & & & & \\
\to &\Xi^0 K^0 & & &\fa(C-D_1+D_2-E-G_1)&& \\
& & & & & & \\
\to &\Xi^- K^+ & & &-\fa(A+D'_2+F)&& \\
& & & & & & \\
\to &p K^{-}  & &&\fa(D_1+E-G_1+G_2)& & \\
& & & & & & \\
\to &n \bar{K}^0 & & &\fa(-C+D_1-D_2+E-G_1) & & \\
& & & & & & \\
\hline
\end{array}
\]
\pagebreak
\begin{center}
 TABLE III. Cabibbo-doubly-suppressed decay amplitudes.
\end{center}
\[ \begin{array}{rllllll}
\hline  \hline
& & & & & &  \\
\mbox{Decay}& \mbox{modes} & & & \qquad \qquad
\mbox{Amplitudes} \quad & & \\
& & & & & &  \\
\hline
& & & & & & \\
\lamc \to & p K^0  & &&\fa(B-C) &  & \\
& & & & & &  \\
\to & n K^+   & & &\fa(A+C)&  & \\
& & & & & & \\
\xia \to &\Sigma^0 K^+  & & &-\sqe(A+D'_2)&&\\
& & & & & &  \\
\to &\Sigma^+ K^0  & & &\fa(B+D_2)&&\\
& & & & & &  \\
\to &\Lambda K^+ & & &\sqf(A+2C+2D'_1-D'_2-2E)  & & \\
& & & & & &  \\
\to & n \pi^+ & & &\fa(-D'_1+E) &&\\
& & & & & & \\
\to & p \pi^0  & & &\sqe(D_2-D'_1+D'_2+E) &&\\
& & & & & &  \\
\to & p \eta_1 & & &\sqd(C-D_2-D'_1+D'_2+E-3H)  & & \\
& & & & & &  \\
\to & p \eta_8  & & &-\sqf(2C+D_2+D'_1-D'_2-E)  & & \\
& & & & & &  \\
\xi0a \to & \Lambda K^0  & & &
 -\frac{1}{2\sqrt{6}}(B-2C+2D_1-D_2+2E) & & \\
& & & & & &  \\
\to &\Sigma^0 K^0  & & &-\sqe(B+D_2) &&\\
& & & & & &  \\
\to &\Sigma^- K^+  & & &-\fa(A+D'_2)& &  \\
& & & & & &  \\
\to & p \pi^-  & & &\fa(D_1+E)& &  \\
& & & & & &  \\
\to & n \pi^0 & & &-\sqe(D_1-D_2-D'_2+E) & & \\
& & & & & &  \\
\to & n \eta_1 & & &\sqd(C+D_1-D_2+D'_2+E-3H)  & & \\
& & & & & &  \\
\to & n \eta_8  & & & -\sqf(2C-D_1+D_2-D'_2-E)  & & \\
& & & & & & \\
\hline
\end{array}
\]

\newpage
\pagestyle{empty}
\setlength{\unitlength}{0.55pt}
\begin{picture}(820,1150)
\put(5,60){
\begin{picture}(810,1090)
\put(0,840){
\begin{picture}(250,250)
\put(125,25){\makebox(0,0){($A$)}}
\put(30,70){\vector(1,0){95}}
\put(125,70){\line(1,0){95}}
\put(30,110){\vector(1,0){95}}
\put(125,110){\line(1,0){95}}
\put(145,240){\vector(0,-1){25}}
\put(145,215){\line(0,-1){5}}
\put(105,240){\line(0,-1){20}}
\put(105,210){\vector(0,1){10}}
\put(125,210){\oval(40,40)[b]}
\put(30,150){\vector(1,0){50}}
\put(80,150){\vector(1,0){95}}
\put(175,150){\line(1,0){45}}
\put(10,70){\line(1,0){10}}
\put(10,70){\line(0,1){40}}
\put(10,110){\line(1,0){10}}
\put(230,70){\line(1,0){10}}
\put(240,70){\line(0,1){40}}
\put(240,110){\line(-1,0){10}}
\multiput(125,150)(0,8){5}{\line(0,1){6}}
\put(50,165){\makebox(0,0){$c$}}
\put(50,125){\makebox(0,0){$q_{2}$}}
\put(200,125){\makebox(0,0){$q_{2}$}}
\put(50,85){\makebox(0,0){$q_{1}$}} 
\put(200,85){\makebox(0,0){$q_{1}$}}
\put(200,165){\makebox(0,0){$q_{3}$}}
\put(160,225){\makebox(0,0){$\overline{q}_{0}$}}
\put(90,225){\makebox(0,0){$q'_{0}$}}
\end{picture}}

\put(280,840){
\begin{picture}(250,250)
\put(125,25){\makebox(0,0){($B$)}}
\put(50,165){\makebox(0,0){$c$}}
\put(50,125){\makebox(0,0){$q_{2}$}}
\put(200,125){\makebox(0,0){$q_{2}$}}
\put(50,85){\makebox(0,0){$q_{1}$}} 
\put(200,85){\makebox(0,0){$q_{1}$}}
\put(200,165){\makebox(0,0){$q'_{0}$}}
\put(160,225){\makebox(0,0){$\overline{q}_{0}$}}
\put(90,225){\makebox(0,0){$q_{3}$}}
\put(70,150){\line(1,0){35}}
\put(30,150){\vector(1,0){40}}
\put(185,150){\line(1,0){35}}
\put(145,150){\vector(1,0){40}}
\put(105,150){\vector(0,1){50}}
\put(105,200){\line(0,1){40}}
\put(145,240){\vector(0,-1){45}}
\put(145,195){\line(0,-1){45}}
\put(30,110){\vector(1,0){95}}
\put(125,110){\line(1,0){95}}
\put(30,70){\vector(1,0){95}}
\put(125,70){\line(1,0){95}}
\put(10,70){\line(1,0){10}}
\put(10,70){\line(0,1){40}}
\put(10,110){\line(1,0){10}}
\put(230,70){\line(1,0){10}}
\put(240,70){\line(0,1){40}}
\put(240,110){\line(-1,0){10}}
\multiput(105,150)(8,0){5}{\line(1,0){6}}
\end{picture}}

\put(560,840){
\begin{picture}(250,250)
\put(125,25){\makebox(0,0){($C$)}}
\put(50,165){\makebox(0,0){$q_2$}}
\put(50,125){\makebox(0,0){$c$}}
\put(200,125){\makebox(0,0){$q_{3}$}}
\put(50,85){\makebox(0,0){$q_{1}$}} 
\put(200,85){\makebox(0,0){$q_{1}$}}
\put(200,165){\makebox(0,0){$q'_{0}$}}
\put(160,225){\makebox(0,0){$\overline{q}_{0}$}}
\put(90,225){\makebox(0,0){$q_{2}$}}
\put(70,150){\line(1,0){35}}
\put(30,150){\vector(1,0){40}}
\put(185,150){\line(1,0){35}}
\put(145,150){\vector(1,0){40}}
\put(105,150){\vector(0,1){50}}
\put(105,200){\line(0,1){40}}
\put(145,240){\vector(0,-1){45}}
\put(145,195){\line(0,-1){45}}
\put(30,110){\vector(1,0){65}}
\put(95,110){\vector(1,0){90}}
\put(185,110){\line(1,0){35}}
\put(30,70){\vector(1,0){95}}
\put(125,70){\line(1,0){95}}
\put(10,70){\line(1,0){10}}
\put(10,70){\line(0,1){80}}
\put(10,150){\line(1,0){10}}
\put(230,110){\line(1,0){10}}
\put(240,110){\line(0,1){40}}
\put(240,150){\line(-1,0){10}}
\multiput(145,110)(0,8){5}{\line(0,1){6}}
\end{picture}}
 
\put(0,560){
\begin{picture}(250,250)
\put(125,25){\makebox(0,0){($D_1$)}}
\put(50,165){\makebox(0,0){$c$}}
\put(50,125){\makebox(0,0){$q_{2}$}}
\put(200,125){\makebox(0,0){$q'_{2}$}}
\put(50,85){\makebox(0,0){$q_{1}$}} 
\put(200,85){\makebox(0,0){$q_{1}$}}
\put(200,165){\makebox(0,0){$q_{0}$}}
\put(160,225){\makebox(0,0){$\overline{q}_{0}$}}
\put(90,225){\makebox(0,0){$q_{3}$}}
\put(70,150){\line(1,0){35}}
\put(30,150){\vector(1,0){40}}
\put(185,150){\line(1,0){35}}
\put(145,150){\vector(1,0){40}}
\put(105,150){\vector(0,1){50}}
\put(105,200){\line(0,1){40}}
\put(145,240){\vector(0,-1){45}}
\put(145,195){\line(0,-1){45}}
\put(30,110){\vector(1,0){40}}
\put(70,110){\vector(1,0){90}}
\put(160,110){\line(1,0){60}}
\put(30,70){\vector(1,0){95}}
\put(125,70){\line(1,0){95}}
\put(10,70){\line(1,0){10}}
\put(10,70){\line(0,1){40}}
\put(10,110){\line(1,0){10}}
\put(230,70){\line(1,0){10}}
\put(240,70){\line(0,1){40}}
\put(240,110){\line(-1,0){10}}
\multiput(105,110)(0,8){5}{\line(0,1){6}}
\end{picture}}

\put(280,560){
\begin{picture}(250,250)
\put(125,25){\makebox(0,0){($D_2$)}}
\put(50,165){\makebox(0,0){$c$}}
\put(50,125){\makebox(0,0){$q_{2}$}}
\put(200,125){\makebox(0,0){$q'_{2}$}}
\put(50,85){\makebox(0,0){$q_{1}$}} 
\put(200,85){\makebox(0,0){$q_{1}$}}
\put(200,165){\makebox(0,0){$q_{0}$}}
\put(160,225){\makebox(0,0){$\overline{q}_{0}$}}
\put(90,225){\makebox(0,0){$q_{3}$}}
\put(70,150){\line(1,0){35}}
\put(30,150){\vector(1,0){40}}
\put(185,150){\line(1,0){35}}
\put(145,150){\vector(1,0){40}}
\put(105,150){\vector(0,1){50}}
\put(105,200){\line(0,1){40}}
\put(145,240){\vector(0,-1){45}}
\put(145,195){\line(0,-1){45}}
\put(30,110){\vector(1,0){40}}
\put(70,110){\vector(1,0){90}}
\put(160,110){\line(1,0){60}}
\put(30,70){\vector(1,0){95}}
\put(125,70){\line(1,0){95}}
\put(10,70){\line(1,0){10}}
\put(10,70){\line(0,1){40}}
\put(10,110){\line(1,0){10}}
\put(230,110){\line(1,0){10}}
\put(240,110){\line(0,1){40}}
\put(240,150){\line(-1,0){10}}
\multiput(105,110)(0,8){5}{\line(0,1){6}}
\end{picture}}

\put(560,560){
\begin{picture}(250,250)
\put(125,25){\makebox(0,0){($D'_1$)}}
\put(50,165){\makebox(0,0){$q_2$}}
\put(50,125){\makebox(0,0){$c$}}
\put(200,125){\makebox(0,0){$q_{3}$}}
\put(50,85){\makebox(0,0){$q_{1}$}} 
\put(200,85){\makebox(0,0){$q_{1}$}}
\put(200,165){\makebox(0,0){$q_{0}$}}
\put(160,225){\makebox(0,0){$\overline{q}_{0}$}}
\put(90,225){\makebox(0,0){$q'_{2}$}}
\put(70,150){\line(1,0){35}}
\put(30,150){\vector(1,0){40}}
\put(185,150){\line(1,0){35}}
\put(145,150){\vector(1,0){40}}
\put(105,150){\vector(0,1){50}}
\put(105,200){\line(0,1){40}}
\put(145,240){\vector(0,-1){45}}
\put(145,195){\line(0,-1){45}}
\put(30,110){\vector(1,0){40}}
\put(70,110){\vector(1,0){90}}
\put(160,110){\line(1,0){60}}
\put(30,70){\vector(1,0){100}}
\put(130,70){\line(1,0){90}}
\put(10,70){\line(1,0){10}}
\put(10,70){\line(0,1){80}}
\put(10,150){\line(1,0){10}}
\put(230,70){\line(1,0){10}}
\put(240,70){\line(0,1){40}}
\put(240,110){\line(-1,0){10}}
\multiput(105,110)(0,8){5}{\line(0,1){6}}
\end{picture}}

\put(0,280){
\begin{picture}(250,250)
\put(125,25){\makebox(0,0){($D'_2$)}}
\put(50,165){\makebox(0,0){$q_2$}}
\put(50,125){\makebox(0,0){$c$}}
\put(200,125){\makebox(0,0){$q_{3}$}}
\put(50,85){\makebox(0,0){$q_{1}$}} 
\put(200,85){\makebox(0,0){$q_{1}$}}
\put(200,165){\makebox(0,0){$q_{0}$}}
\put(160,225){\makebox(0,0){$\overline{q}_{0}$}}
\put(90,225){\makebox(0,0){$q'_{2}$}}
\put(70,150){\line(1,0){35}}
\put(30,150){\vector(1,0){40}}
\put(185,150){\line(1,0){35}}
\put(145,150){\vector(1,0){40}}
\put(105,150){\vector(0,1){50}}
\put(105,200){\line(0,1){40}}
\put(145,240){\vector(0,-1){45}}
\put(145,195){\line(0,-1){45}}
\put(30,110){\vector(1,0){40}}
\put(70,110){\vector(1,0){90}}
\put(160,110){\line(1,0){60}}
\put(30,70){\vector(1,0){95}}
\put(125,70){\line(1,0){95}}
\put(10,70){\line(1,0){10}}
\put(10,70){\line(0,1){80}}
\put(10,150){\line(1,0){10}}
\put(230,110){\line(1,0){10}}
\put(240,110){\line(0,1){40}}
\put(240,150){\line(-1,0){10}}
\multiput(105,110)(0,8){5}{\line(0,1){6}}
\end{picture}}

\put(280,280){
\begin{picture}(250,250)
\put(125,25){\makebox(0,0){($E$)}}
\put(50,165){\makebox(0,0){$q_2$}}
\put(50,125){\makebox(0,0){$c$}}
\put(200,125){\makebox(0,0){$q_{3}$}}
\put(50,85){\makebox(0,0){$q_{1}$}} 
\put(200,85){\makebox(0,0){$q'_{1}$}}
\put(200,165){\makebox(0,0){$q_{0}$}}
\put(160,225){\makebox(0,0){$\overline{q}_{0}$}}
\put(90,225){\makebox(0,0){$q_{2}$}}
\put(70,150){\line(1,0){35}}
\put(30,150){\vector(1,0){40}}
\put(185,150){\line(1,0){35}}
\put(145,150){\vector(1,0){40}}
\put(105,150){\vector(0,1){50}}
\put(105,200){\line(0,1){40}}
\put(145,240){\vector(0,-1){45}}
\put(145,195){\line(0,-1){45}}
\put(30,110){\vector(1,0){50}}
\put(80,110){\vector(1,0){95}}
\put(175,110){\line(1,0){45}}
\put(30,70){\vector(1,0){50}}
\put(80,70){\vector(1,0){95}}
\put(175,70){\line(1,0){45}}
\put(10,70){\line(1,0){10}}
\put(10,70){\line(0,1){80}}
\put(10,150){\line(1,0){10}}
\put(230,70){\line(1,0){10}}
\put(240,70){\line(0,1){40}}
\put(240,110){\line(-1,0){10}}
\multiput(125,70)(0,8){5}{\line(0,1){6}}
\end{picture}}

\put(560,280){
\begin{picture}(250,250)
\put(125,25){\makebox(0,0){($F$)}}
\put(50,165){\makebox(0,0){$c$}}
\put(50,125){\makebox(0,0){$q_{2}$}}
\put(200,125){\makebox(0,0){$q_{2}$}}
\put(50,85){\makebox(0,0){$q_{1}$}} 
\put(200,85){\makebox(0,0){$q_{1}$}}
\put(200,165){\makebox(0,0){$q_{0}$}}
\put(195,225){\makebox(0,0){$\overline{q}_{0}$}}
\put(125,225){\makebox(0,0){$q_{3}$}}
\put(55,150){\line(1,0){15}}
\put(30,150){\vector(1,0){25}}
\put(90,150){\oval(40,40)[b]}
\put(110,150){\vector(1,0){20}}
\put(130,150){\line(1,0){10}}
\put(205,150){\line(1,0){15}}
\put(180,150){\vector(1,0){25}}
\put(140,150){\vector(0,1){50}}
\put(140,200){\line(0,1){40}}
\put(180,240){\vector(0,-1){45}}
\put(180,195){\line(0,-1){45}}
\put(30,110){\vector(1,0){95}}
\put(125,110){\line(1,0){95}}
\put(30,70){\vector(1,0){95}}
\put(125,70){\line(1,0){95}}
\put(10,70){\line(1,0){10}}
\put(10,70){\line(0,1){40}}
\put(10,110){\line(1,0){10}}
\put(230,70){\line(1,0){10}}
\put(240,70){\line(0,1){40}}
\put(240,110){\line(-1,0){10}}
\multiput(70,150)(8,0){5}{\line(1,0){6}}
\end{picture}}

\put(0,0){
\begin{picture}(250,250)
\put(125,25){\makebox(0,0){($G_1$)}}
\put(50,165){\makebox(0,0){$q_2$}}
\put(50,125){\makebox(0,0){$q_{1}$}}
\put(200,125){\makebox(0,0){$q_{1}$}}
\put(50,85){\makebox(0,0){$c$}} 
\put(200,85){\makebox(0,0){$q_{3}$}}
\put(200,165){\makebox(0,0){$q_{0}$}}
\put(160,225){\makebox(0,0){$\overline{q}_{0}$}}
\put(90,225){\makebox(0,0){$q_{2}$}}
\put(70,150){\line(1,0){35}}
\put(30,150){\vector(1,0){40}}
\put(185,150){\line(1,0){35}}
\put(145,150){\vector(1,0){40}}
\put(105,150){\vector(0,1){50}}
\put(105,200){\line(0,1){40}}
\put(145,240){\vector(0,-1){45}}
\put(145,195){\line(0,-1){45}}
\put(30,110){\vector(1,0){100}}
\put(130,110){\line(1,0){90}}
\put(30,70){\vector(1,0){40}}
\put(70,70){\line(1,0){35}}
\put(125,70){\oval(40,40)[t]}
\put(145,70){\vector(1,0){40}}
\put(175,70){\line(1,0){35}}
\put(10,110){\line(1,0){10}}
\put(10,110){\line(0,1){40}}
\put(10,150){\line(1,0){10}}
\put(230,70){\line(1,0){10}}
\put(240,70){\line(0,1){40}}
\put(240,110){\line(-1,0){10}}
\multiput(105,70)(8,0){5}{\line(1,0){6}}
\end{picture}}

\put(280,0){
\begin{picture}(250,250)
\put(125,25){\makebox(0,0){($G_2$)}}
\put(50,165){\makebox(0,0){$q_2$}}
\put(50,125){\makebox(0,0){$q_{1}$}}
\put(200,125){\makebox(0,0){$q_{1}$}}
\put(50,85){\makebox(0,0){$c$}} 
\put(200,85){\makebox(0,0){$q_{3}$}}
\put(200,165){\makebox(0,0){$q_{0}$}}
\put(160,225){\makebox(0,0){$\overline{q}_{0}$}}
\put(90,225){\makebox(0,0){$q_{2}$}}
\put(70,150){\line(1,0){35}}
\put(30,150){\vector(1,0){40}}
\put(185,150){\line(1,0){35}}
\put(145,150){\vector(1,0){40}}
\put(105,150){\vector(0,1){50}}
\put(105,200){\line(0,1){40}}
\put(145,240){\vector(0,-1){45}}
\put(145,195){\line(0,-1){45}}
\put(30,110){\vector(1,0){100}}
\put(130,110){\line(1,0){90}}
\put(30,70){\vector(1,0){40}}
\put(70,70){\line(1,0){35}}
\put(125,70){\oval(40,40)[t]}
\put(145,70){\vector(1,0){40}}
\put(175,70){\line(1,0){35}}
\put(10,110){\line(1,0){10}}
\put(10,110){\line(0,1){40}}
\put(10,150){\line(1,0){10}}
\put(230,110){\line(1,0){10}}
\put(240,110){\line(0,1){40}}
\put(240,150){\line(-1,0){10}}
\multiput(105,70)(8,0){5}{\line(1,0){6}}
\end{picture}}

\put(560,0){
\begin{picture}(250,250)
\put(125,25){\makebox(0,0){($H$)}}
\put(30,70){\vector(1,0){95}}
\put(125,70){\line(1,0){95}}
\put(30,110){\vector(1,0){50}}
\put(80,110){\vector(1,0){95}}
\put(175,110){\line(1,0){45}}
\put(145,240){\vector(0,-1){25}}
\put(145,215){\line(0,-1){5}}
\put(105,240){\line(0,-1){20}}
\put(105,210){\vector(0,1){10}}
\put(125,210){\oval(40,40)[b]}
\put(30,150){\vector(1,0){50}}
\put(80,150){\vector(1,0){95}}
\put(175,150){\line(1,0){45}}
\put(10,70){\line(1,0){10}}
\put(10,70){\line(0,1){40}}
\put(10,110){\line(1,0){10}}
\put(230,110){\line(1,0){10}}
\put(240,110){\line(0,1){40}}
\put(240,150){\line(-1,0){10}}
\multiput(125,110)(0,8){5}{\line(0,1){6}}
\put(50,165){\makebox(0,0){$c$}}
\put(50,125){\makebox(0,0){$q_{2}$}}
\put(200,125){\makebox(0,0){$q'_{2}$}}
\put(50,85){\makebox(0,0){$q_{1}$}} 
\put(200,85){\makebox(0,0){$q_{1}$}}
\put(200,165){\makebox(0,0){$q_{3}$}}
\put(160,225){\makebox(0,0){$\overline{q}_{0}$}}
\put(90,225){\makebox(0,0){$q_{0}$}}
\end{picture}}
\end{picture}}
\put(405,10){\makebox(0,0)[b]{Fig. 1.
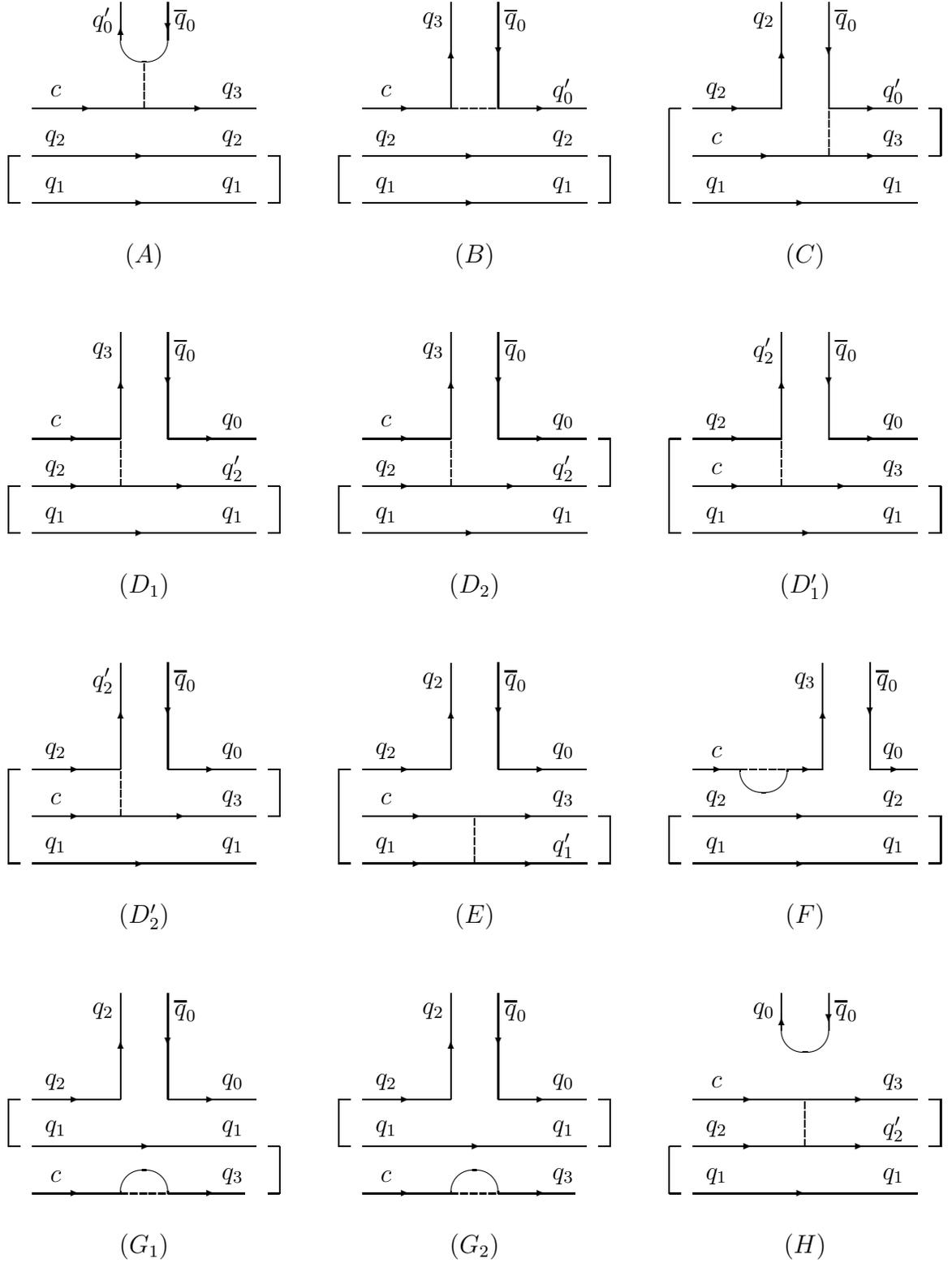
 Quark diagrams in antisymmetric bases}}
\end{picture}

\newpage
\pagestyle{empty}

\setlength{\unitlength}{0.55pt}
\begin{picture}(820,1150)
\put(0,60){
\begin{picture}(820,1090)

\put(0,900){
\begin{picture}(180,180)
\put(30,60){\vector(1,0){65}}
\put(95,60){\line(1,0){55}}
\put(30,90){\vector(1,0){65}}
\put(95,90){\line(1,0){55}}
\put(105,180){\vector(0,-1){15}}
\put(75,180){\line(0,-1){10}}
\put(75,165){\vector(0,1){5}}
\put(90,165){\oval(30,30)[b]}
\put(30,120){\vector(1,0){35}}
\put(65,120){\vector(1,0){55}}
\put(120,120){\line(1,0){30}}
\put(10,60){\line(1,0){10}}
\put(10,60){\line(0,1){30}}
\put(10,90){\line(1,0){10}}
\put(160,105){\oval(20,30)[r]}
\multiput(90,120)(0,6){5}{\line(0,1){4}}
\put(40,135){\makebox(0,0){$c$}}
\put(40,75){\makebox(0,0){$q_{2}$}}
\put(140,75){\makebox(0,0){$q_{2}$}}
\put(40,45){\makebox(0,0){$q_{1}$}} 
\put(140,45){\makebox(0,0){$q_{1}$}}
\put(140,135){\makebox(0,0){$q_{3}$}}
\put(120,165){\makebox(0,0){$\overline{q}_{0}$}}
\put(60,165){\makebox(0,0){$q'_{0}$}}
\end{picture}}

\put(210,900){
\begin{picture}(180,180)
\put(30,60){\vector(1,0){65}}
\put(95,60){\line(1,0){55}}
\put(30,90){\vector(1,0){65}}
\put(95,90){\line(1,0){55}}
\put(105,180){\vector(0,-1){15}}
\put(75,180){\line(0,-1){10}}
\put(75,165){\vector(0,1){5}}
\put(90,165){\oval(30,30)[b]}
\put(30,120){\vector(1,0){35}}
\put(65,120){\vector(1,0){55}}
\put(120,120){\line(1,0){30}}
\put(10,60){\line(1,0){10}}
\put(10,60){\line(0,1){30}}
\put(10,90){\line(1,0){10}}
\put(160,90){\oval(20,60)[r]}
\multiput(90,120)(0,6){5}{\line(0,1){4}}
\put(40,135){\makebox(0,0){$c$}}
\put(40,75){\makebox(0,0){$q_{2}$}}
\put(140,75){\makebox(0,0){$q_{2}$}}
\put(40,45){\makebox(0,0){$q_{1}$}} 
\put(140,45){\makebox(0,0){$q_{1}$}}
\put(140,135){\makebox(0,0){$q_{3}$}}
\put(120,165){\makebox(0,0){$\overline{q}_{0}$}}
\put(60,165){\makebox(0,0){$q'_{0}$}}
\end{picture}}

\put(420,900){
\begin{picture}(180,180)
\put(30,60){\vector(1,0){65}}
\put(95,60){\line(1,0){55}}
\put(30,90){\vector(1,0){65}}
\put(95,90){\line(1,0){55}}
\put(105,180){\vector(0,-1){30}}
\put(105,150){\line(0,-1){30}}
\put(75,180){\line(0,-1){25}}
\put(75,120){\vector(0,1){35}}
\put(30,120){\vector(1,0){25}}
\put(55,120){\line(1,0){20}}
\put(105,120){\vector(1,0){25}}
\put(130,120){\line(1,0){20}}
\put(10,60){\line(1,0){10}}
\put(10,60){\line(0,1){30}}
\put(10,90){\line(1,0){10}}
\put(160,105){\oval(20,30)[r]}
\multiput(75,120)(6,0){5}{\line(1,0){4}}
\put(40,135){\makebox(0,0){$c$}}
\put(40,75){\makebox(0,0){$q_{2}$}}
\put(140,75){\makebox(0,0){$q_{2}$}}
\put(40,45){\makebox(0,0){$q_{1}$}} 
\put(140,45){\makebox(0,0){$q_{1}$}}
\put(140,135){\makebox(0,0){$q'_{0}$}}
\put(120,165){\makebox(0,0){$\overline{q}_{0}$}}
\put(60,165){\makebox(0,0){$q_{3}$}}
\end{picture}}

\put(630,900){
\begin{picture}(180,180)
\put(30,60){\vector(1,0){65}}
\put(95,60){\line(1,0){55}}
\put(30,90){\vector(1,0){65}}
\put(95,90){\line(1,0){55}}
\put(105,180){\vector(0,-1){30}}
\put(105,150){\line(0,-1){30}}
\put(75,180){\line(0,-1){25}}
\put(75,120){\vector(0,1){35}}
\put(30,120){\vector(1,0){25}}
\put(55,120){\line(1,0){20}}
\put(105,120){\vector(1,0){25}}
\put(130,120){\line(1,0){20}}
\put(10,60){\line(1,0){10}}
\put(10,60){\line(0,1){30}}
\put(10,90){\line(1,0){10}}
\put(160,90){\oval(20,60)[r]}
\multiput(75,120)(6,0){5}{\line(1,0){4}}
\put(40,135){\makebox(0,0){$c$}}
\put(40,75){\makebox(0,0){$q_{2}$}}
\put(140,75){\makebox(0,0){$q_{2}$}}
\put(40,45){\makebox(0,0){$q_{1}$}} 
\put(140,45){\makebox(0,0){$q_{1}$}}
\put(140,135){\makebox(0,0){$q'_{0}$}}
\put(120,165){\makebox(0,0){$\overline{q}_{0}$}}
\put(60,165){\makebox(0,0){$q_{3}$}}
\end{picture}}

\put(0,675){
\begin{picture}(180,180)
\put(30,60){\vector(1,0){65}}
\put(95,60){\line(1,0){55}}
\put(30,90){\vector(1,0){45}}
\put(75,90){\vector(1,0){55}}
\put(130,90){\line(1,0){20}}
\put(105,180){\vector(0,-1){30}}
\put(105,150){\line(0,-1){30}}
\put(75,180){\line(0,-1){25}}
\put(75,120){\vector(0,1){35}}
\put(30,120){\vector(1,0){25}}
\put(55,120){\line(1,0){20}}
\put(105,120){\vector(1,0){25}}
\put(130,120){\line(1,0){20}}
\put(10,60){\line(1,0){10}}
\put(10,60){\line(0,1){60}}
\put(10,120){\line(1,0){10}}
\put(160,75){\oval(20,30)[r]}
\multiput(105,120)(0,-6){5}{\line(0,-1){4}}
\put(40,135){\makebox(0,0){$q_2$}}
\put(40,75){\makebox(0,0){$c$}}
\put(140,75){\makebox(0,0){$q_{3}$}}
\put(40,45){\makebox(0,0){$q_{1}$}} 
\put(140,45){\makebox(0,0){$q_{1}$}}
\put(140,135){\makebox(0,0){$q'_{0}$}}
\put(120,165){\makebox(0,0){$\overline{q}_{0}$}}
\put(60,165){\makebox(0,0){$q_{2}$}}
\end{picture}}

\put(210,675){
\begin{picture}(180,180)
\put(30,60){\vector(1,0){65}}
\put(95,60){\line(1,0){55}}
\put(30,90){\vector(1,0){45}}
\put(75,90){\vector(1,0){55}}
\put(130,90){\line(1,0){20}}
\put(105,180){\vector(0,-1){30}}
\put(105,150){\line(0,-1){30}}
\put(75,180){\line(0,-1){25}}
\put(75,120){\vector(0,1){35}}
\put(30,120){\vector(1,0){25}}
\put(55,120){\line(1,0){20}}
\put(105,120){\vector(1,0){25}}
\put(130,120){\line(1,0){20}}
\put(10,60){\line(1,0){10}}
\put(10,60){\line(0,1){60}}
\put(10,120){\line(1,0){10}}
\put(160,90){\oval(20,60)[r]}
\multiput(105,120)(0,-6){5}{\line(0,-1){4}}
\put(40,135){\makebox(0,0){$q_2$}}
\put(40,75){\makebox(0,0){$c$}}
\put(140,75){\makebox(0,0){$q_{3}$}}
\put(40,45){\makebox(0,0){$q_{1}$}} 
\put(140,45){\makebox(0,0){$q_{1}$}}
\put(140,135){\makebox(0,0){$q'_{0}$}}
\put(120,165){\makebox(0,0){$\overline{q}_{0}$}}
\put(60,165){\makebox(0,0){$q_{2}$}}
\end{picture}} 

\put(420,675){
\begin{picture}(180,180)
\put(90,15){\makebox(0,0){($D_{S1}$)}}
\put(30,60){\vector(1,0){65}}
\put(95,60){\line(1,0){55}}
\put(30,90){\vector(1,0){25}}
\put(55,90){\vector(1,0){55}}
\put(110,90){\line(1,0){40}}
\put(105,180){\vector(0,-1){30}}
\put(105,150){\line(0,-1){30}}
\put(75,180){\line(0,-1){25}}
\put(75,120){\vector(0,1){35}}
\put(30,120){\vector(1,0){25}}
\put(55,120){\line(1,0){20}}
\put(105,120){\vector(1,0){25}}
\put(130,120){\line(1,0){20}}
\put(10,60){\line(1,0){10}}
\put(10,60){\line(0,1){30}}
\put(10,90){\line(1,0){10}}
\put(160,105){\oval(20,30)[r]}
\multiput(75,120)(0,-6){5}{\line(0,-1){4}}
\put(40,135){\makebox(0,0){$c$}}
\put(40,75){\makebox(0,0){$q_2$}}
\put(140,75){\makebox(0,0){$q'_{2}$}}
\put(40,45){\makebox(0,0){$q_{1}$}} 
\put(140,45){\makebox(0,0){$q_{1}$}}
\put(140,135){\makebox(0,0){$q'_{0}$}}
\put(120,165){\makebox(0,0){$\overline{q}_{0}$}}
\put(60,165){\makebox(0,0){$q_{3}$}}
\end{picture}}

\put(630,675){
\begin{picture}(180,180)
\put(90,15){\makebox(0,0){($D_{S2}$)}}
\put(30,60){\vector(1,0){65}}
\put(95,60){\line(1,0){55}}
\put(30,90){\vector(1,0){25}}
\put(55,90){\vector(1,0){55}}
\put(110,90){\line(1,0){40}}
\put(105,180){\vector(0,-1){30}}
\put(105,150){\line(0,-1){30}}
\put(75,180){\line(0,-1){25}}
\put(75,120){\vector(0,1){35}}
\put(30,120){\vector(1,0){25}}
\put(55,120){\line(1,0){20}}
\put(105,120){\vector(1,0){25}}
\put(130,120){\line(1,0){20}}
\put(10,60){\line(1,0){10}}
\put(10,60){\line(0,1){30}}
\put(10,90){\line(1,0){10}}
\put(160,90){\oval(20,60)[r]}
\multiput(75,120)(0,-6){5}{\line(0,-1){4}}
\put(40,135){\makebox(0,0){$c$}}
\put(40,75){\makebox(0,0){$q_2$}}
\put(140,75){\makebox(0,0){$q'_{2}$}}
\put(40,45){\makebox(0,0){$q_{1}$}} 
\put(140,45){\makebox(0,0){$q_{1}$}}
\put(140,135){\makebox(0,0){$q'_{0}$}}
\put(120,165){\makebox(0,0){$\overline{q}_{0}$}}
\put(60,165){\makebox(0,0){$q_{3}$}}
\end{picture}}

\put(0,450){
\begin{picture}(180,180)
\put(90,15){\makebox(0,0){($D'_{S1}$)}}
\put(30,60){\vector(1,0){65}}
\put(95,60){\line(1,0){55}}
\put(30,90){\vector(1,0){25}}
\put(55,90){\vector(1,0){55}}
\put(110,90){\line(1,0){40}}
\put(105,180){\vector(0,-1){30}}
\put(105,150){\line(0,-1){30}}
\put(75,180){\line(0,-1){25}}
\put(75,120){\vector(0,1){35}}
\put(30,120){\vector(1,0){25}}
\put(55,120){\line(1,0){20}}
\put(105,120){\vector(1,0){25}}
\put(130,120){\line(1,0){20}}
\put(10,60){\line(1,0){10}}
\put(10,60){\line(0,1){60}}
\put(10,120){\line(1,0){10}}
\put(160,105){\oval(20,30)[r]}
\multiput(75,120)(0,-6){5}{\line(0,-1){4}}
\put(40,135){\makebox(0,0){$q_2$}}
\put(40,75){\makebox(0,0){$c$}}
\put(140,75){\makebox(0,0){$q_{3}$}}
\put(40,45){\makebox(0,0){$q_{1}$}} 
\put(140,45){\makebox(0,0){$q_{1}$}}
\put(140,135){\makebox(0,0){$q_{0}$}}
\put(120,165){\makebox(0,0){$\overline{q}_{0}$}}
\put(60,165){\makebox(0,0){$q'_{2}$}}
\end{picture}}

\put(210,450){
\begin{picture}(180,180)
\put(90,15){\makebox(0,0){($D'_{S2}$)}}
\put(30,60){\vector(1,0){65}}
\put(95,60){\line(1,0){55}}
\put(30,90){\vector(1,0){25}}
\put(55,90){\vector(1,0){55}}
\put(110,90){\line(1,0){40}}
\put(105,180){\vector(0,-1){30}}
\put(105,150){\line(0,-1){30}}
\put(75,180){\line(0,-1){25}}
\put(75,120){\vector(0,1){35}}
\put(30,120){\vector(1,0){25}}
\put(55,120){\line(1,0){20}}
\put(105,120){\vector(1,0){25}}
\put(130,120){\line(1,0){20}}
\put(10,60){\line(1,0){10}}
\put(10,60){\line(0,1){60}}
\put(10,120){\line(1,0){10}}
\put(160,90){\oval(20,60)[r]}
\multiput(75,120)(0,-6){5}{\line(0,-1){4}}
\put(40,135){\makebox(0,0){$q_2$}}
\put(40,75){\makebox(0,0){$c$}}
\put(140,75){\makebox(0,0){$q_{3}$}}
\put(40,45){\makebox(0,0){$q_{1}$}} 
\put(140,45){\makebox(0,0){$q_{1}$}}
\put(140,135){\makebox(0,0){$q_{0}$}}
\put(120,165){\makebox(0,0){$\overline{q}_{0}$}}
\put(60,165){\makebox(0,0){$q'_{2}$}}
\end{picture}}

\put(420,450){
\begin{picture}(180,180)
\put(30,60){\vector(1,0){35}}
\put(65,60){\vector(1,0){60}}
\put(125,60){\line(1,0){30}}
\put(30,90){\vector(1,0){35}}
\put(65,90){\vector(1,0){60}}
\put(125,90){\line(1,0){30}}
\put(105,180){\vector(0,-1){30}}
\put(105,150){\line(0,-1){30}}
\put(75,180){\line(0,-1){25}}
\put(75,120){\vector(0,1){35}}
\put(30,120){\vector(1,0){25}}
\put(55,120){\line(1,0){20}}
\put(105,120){\vector(1,0){25}}
\put(130,120){\line(1,0){20}}
\put(10,60){\line(1,0){10}}
\put(10,60){\line(0,1){60}}
\put(10,120){\line(1,0){10}}
\put(160,105){\oval(20,30)[r]}
\multiput(90,60)(0,6){5}{\line(0,1){4}}
\put(40,135){\makebox(0,0){$q_2$}}
\put(40,75){\makebox(0,0){$c$}}
\put(140,75){\makebox(0,0){$q_{3}$}}
\put(40,45){\makebox(0,0){$q_{1}$}} 
\put(140,45){\makebox(0,0){$q'_{1}$}}
\put(140,135){\makebox(0,0){$q_{0}$}}
\put(120,165){\makebox(0,0){$\overline{q}_{0}$}}
\put(60,165){\makebox(0,0){$q_{2}$}}
\end{picture}}

\put(630,450){
\begin{picture}(180,180)
\put(30,60){\vector(1,0){35}}
\put(65,60){\vector(1,0){60}}
\put(125,60){\line(1,0){30}}
\put(30,90){\vector(1,0){35}}
\put(65,90){\vector(1,0){60}}
\put(125,90){\line(1,0){30}}
\put(105,180){\vector(0,-1){30}}
\put(105,150){\line(0,-1){30}}
\put(75,180){\line(0,-1){25}}
\put(75,120){\vector(0,1){35}}
\put(30,120){\vector(1,0){25}}
\put(55,120){\line(1,0){20}}
\put(105,120){\vector(1,0){25}}
\put(130,120){\line(1,0){20}}
\put(10,60){\line(1,0){10}}
\put(10,60){\line(0,1){60}}
\put(10,120){\line(1,0){10}}
\put(160,90){\oval(20,60)[r]}
\multiput(90,60)(0,6){5}{\line(0,1){4}}
\put(40,135){\makebox(0,0){$q_2$}}
\put(40,75){\makebox(0,0){$c$}}
\put(140,75){\makebox(0,0){$q_{3}$}}
\put(40,45){\makebox(0,0){$q_{1}$}} 
\put(140,45){\makebox(0,0){$q'_{1}$}}
\put(140,135){\makebox(0,0){$q_{0}$}}
\put(120,165){\makebox(0,0){$\overline{q}_{0}$}}
\put(60,165){\makebox(0,0){$q_{2}$}}
\end{picture}}

\put(0,225){
\begin{picture}(180,180)
\put(30,60){\vector(1,0){65}}
\put(95,60){\line(1,0){55}}
\put(30,90){\vector(1,0){65}}
\put(95,90){\line(1,0){55}}
\put(130,180){\vector(0,-1){30}}
\put(130,150){\line(0,-1){30}}
\put(100,180){\line(0,-1){25}}
\put(100,120){\vector(0,1){35}}
\put(30,120){\vector(1,0){13}}
\put(43,120){\line(1,0){7}}
\put(65,120){\oval(30,30)[b]}
\put(80,120){\vector(1,0){13}}
\put(93,120){\line(1,0){7}}
\put(130,120){\vector(1,0){13}}
\put(143,120){\line(1,0){7}}
\put(10,60){\line(1,0){10}}
\put(10,60){\line(0,1){30}}
\put(10,90){\line(1,0){10}}
\put(160,105){\oval(20,30)[r]}
\multiput(50,120)(6,0){5}{\line(1,0){4}}
\put(40,135){\makebox(0,0){$c$}}
\put(40,75){\makebox(0,0){$q_2$}}
\put(140,75){\makebox(0,0){$q_{2}$}}
\put(40,45){\makebox(0,0){$q_{1}$}} 
\put(140,45){\makebox(0,0){$q_{1}$}}
\put(145,135){\makebox(0,0){$q_{0}$}}
\put(145,165){\makebox(0,0){$\overline{q}_{0}$}}
\put(85,165){\makebox(0,0){$q_{3}$}}
\end{picture}}

\put(210,225){
\begin{picture}(180,180)
\put(30,60){\vector(1,0){65}}
\put(95,60){\line(1,0){55}}
\put(30,90){\vector(1,0){65}}
\put(95,90){\line(1,0){55}}
\put(130,180){\vector(0,-1){30}}
\put(130,150){\line(0,-1){30}}
\put(100,180){\line(0,-1){25}}
\put(100,120){\vector(0,1){35}}
\put(30,120){\vector(1,0){13}}
\put(43,120){\line(1,0){7}}
\put(65,120){\oval(30,30)[b]}
\put(80,120){\vector(1,0){13}}
\put(93,120){\line(1,0){7}}
\put(130,120){\vector(1,0){13}}
\put(143,120){\line(1,0){7}}
\put(10,60){\line(1,0){10}}
\put(10,60){\line(0,1){30}}
\put(10,90){\line(1,0){10}}
\put(160,90){\oval(20,60)[r]}
\multiput(50,120)(6,0){5}{\line(1,0){4}}
\put(40,135){\makebox(0,0){$c$}}
\put(40,75){\makebox(0,0){$q_2$}}
\put(140,75){\makebox(0,0){$q_{2}$}}
\put(40,45){\makebox(0,0){$q_{1}$}} 
\put(140,45){\makebox(0,0){$q_{1}$}}
\put(145,135){\makebox(0,0){$q_{0}$}}
\put(145,165){\makebox(0,0){$\overline{q}_{0}$}}
\put(85,165){\makebox(0,0){$q_{3}$}}
\end{picture}}

\put(420,225){
\begin{picture}(180,180)
\put(90,15){\makebox(0,0){($G_{S1}$)}}
\put(30,60){\vector(1,0){25}}
\put(55,60){\line(1,0){20}}
\put(90,60){\oval(30,30)[t]}
\put(105,60){\vector(1,0){25}}
\put(130,60){\line(1,0){20}}
\put(30,90){\vector(1,0){65}}
\put(95,90){\line(1,0){55}}
\put(105,180){\vector(0,-1){30}}
\put(105,150){\line(0,-1){30}}
\put(75,180){\line(0,-1){25}}
\put(75,120){\vector(0,1){35}}
\put(30,120){\vector(1,0){25}}
\put(55,120){\line(1,0){20}}
\put(105,120){\vector(1,0){25}}
\put(130,120){\line(1,0){20}}
\put(10,90){\line(1,0){10}}
\put(10,90){\line(0,1){30}}
\put(10,120){\line(1,0){10}}
\put(160,105){\oval(20,30)[r]}
\multiput(75,60)(6,0){5}{\line(1,0){4}}
\put(40,135){\makebox(0,0){$q_2$}}
\put(40,75){\makebox(0,0){$q_1$}}
\put(140,75){\makebox(0,0){$q_{1}$}}
\put(40,45){\makebox(0,0){$c$}} 
\put(140,45){\makebox(0,0){$q_{3}$}}
\put(140,135){\makebox(0,0){$q_{0}$}}
\put(120,165){\makebox(0,0){$\overline{q}_{0}$}}
\put(60,165){\makebox(0,0){$q_{2}$}}
\end{picture}}

\put(630,225){
\begin{picture}(180,180)
\put(90,15){\makebox(0,0){($G_{S2}$)}}
\put(30,60){\vector(1,0){25}}
\put(55,60){\line(1,0){20}}
\put(90,60){\oval(30,30)[t]}
\put(105,60){\vector(1,0){25}}
\put(130,60){\line(1,0){20}}
\put(30,90){\vector(1,0){65}}
\put(95,90){\line(1,0){55}}
\put(105,180){\vector(0,-1){30}}
\put(105,150){\line(0,-1){30}}
\put(75,180){\line(0,-1){25}}
\put(75,120){\vector(0,1){35}}
\put(30,120){\vector(1,0){25}}
\put(55,120){\line(1,0){20}}
\put(105,120){\vector(1,0){25}}
\put(130,120){\line(1,0){20}}
\put(10,90){\line(1,0){10}}
\put(10,90){\line(0,1){30}}
\put(10,120){\line(1,0){10}}
\put(160,90){\oval(20,60)[r]}
\multiput(75,60)(6,0){5}{\line(1,0){4}}
\put(40,135){\makebox(0,0){$q_2$}}
\put(40,75){\makebox(0,0){$q_1$}}
\put(140,75){\makebox(0,0){$q_{1}$}}
\put(40,45){\makebox(0,0){$c$}} 
\put(140,45){\makebox(0,0){$q_{3}$}}
\put(140,135){\makebox(0,0){$q_{0}$}}
\put(120,165){\makebox(0,0){$\overline{q}_{0}$}}
\put(60,165){\makebox(0,0){$q_{2}$}}
\end{picture}}

\put(0,0){
\begin{picture}(180,180)
\put(30,60){\vector(1,0){65}}
\put(95,60){\line(1,0){55}}
\put(30,90){\vector(1,0){35}}
\put(65,90){\vector(1,0){55}}
\put(120,90){\line(1,0){30}}
\put(105,180){\vector(0,-1){15}}
\put(75,180){\line(0,-1){10}}
\put(75,165){\vector(0,1){5}}
\put(90,165){\oval(30,30)[b]}
\put(30,120){\vector(1,0){35}}
\put(65,120){\vector(1,0){55}}
\put(120,120){\line(1,0){30}}
\put(10,60){\line(1,0){10}}
\put(10,60){\line(0,1){30}}
\put(10,90){\line(1,0){10}}
\put(160,90){\oval(20,60)[r]}
\multiput(90,90)(0,6){5}{\line(0,1){4}}
\put(40,135){\makebox(0,0){$c$}}
\put(40,75){\makebox(0,0){$q_{2}$}}
\put(140,75){\makebox(0,0){$q'_{2}$}}
\put(40,45){\makebox(0,0){$q_{1}$}} 
\put(140,45){\makebox(0,0){$q_{1}$}}
\put(140,135){\makebox(0,0){$q_{3}$}}
\put(120,165){\makebox(0,0){$\overline{q}_{0}$}}
\put(60,165){\makebox(0,0){$q_{0}$}}
\end{picture}}

\put(210,0){
\begin{picture}(180,180)
\put(30,60){\vector(1,0){65}}
\put(95,60){\line(1,0){55}}
\put(30,90){\vector(1,0){35}}
\put(65,90){\vector(1,0){55}}
\put(120,90){\line(1,0){30}}
\put(105,180){\vector(0,-1){15}}
\put(75,180){\line(0,-1){10}}
\put(75,165){\vector(0,1){5}}
\put(90,165){\oval(30,30)[b]}
\put(30,120){\vector(1,0){35}}
\put(65,120){\vector(1,0){55}}
\put(120,120){\line(1,0){30}}
\put(10,60){\line(1,0){10}}
\put(10,60){\line(0,1){30}}
\put(10,90){\line(1,0){10}}
\put(160,75){\oval(20,30)[r]}
\multiput(90,90)(0,6){5}{\line(0,1){4}}
\put(40,135){\makebox(0,0){$c$}}
\put(40,75){\makebox(0,0){$q_{2}$}}
\put(140,75){\makebox(0,0){$q'_{2}$}}
\put(40,45){\makebox(0,0){$q_{1}$}} 
\put(140,45){\makebox(0,0){$q_{1}$}}
\put(140,135){\makebox(0,0){$q_{3}$}}
\put(120,165){\makebox(0,0){$\overline{q}_{0}$}}
\put(60,165){\makebox(0,0){$q_{0}$}}
\end{picture}}
\end{picture}}
\put(210,1065){\makebox(0,0){$-$}}
\put(210,840){\makebox(0,0){$-$}}
\put(210,390){\makebox(0,0){$-$}}
\put(210,165){\makebox(0,0){$-$}}
\put(630,1065){\makebox(0,0){$-$}}
\put(630,615){\makebox(0,0){$-$}}
\put(210,980){\makebox(0,0){$(A_S)$}}
\put(210,755){\makebox(0,0){$(C_S)$}}
\put(210,305){\makebox(0,0){$(F_S)$}}
\put(210,80){\makebox(0,0){$(H_S)$}}
\put(630,980){\makebox(0,0){$(B_S)$}}
\put(630,530){\makebox(0,0){$(E_S)$}}
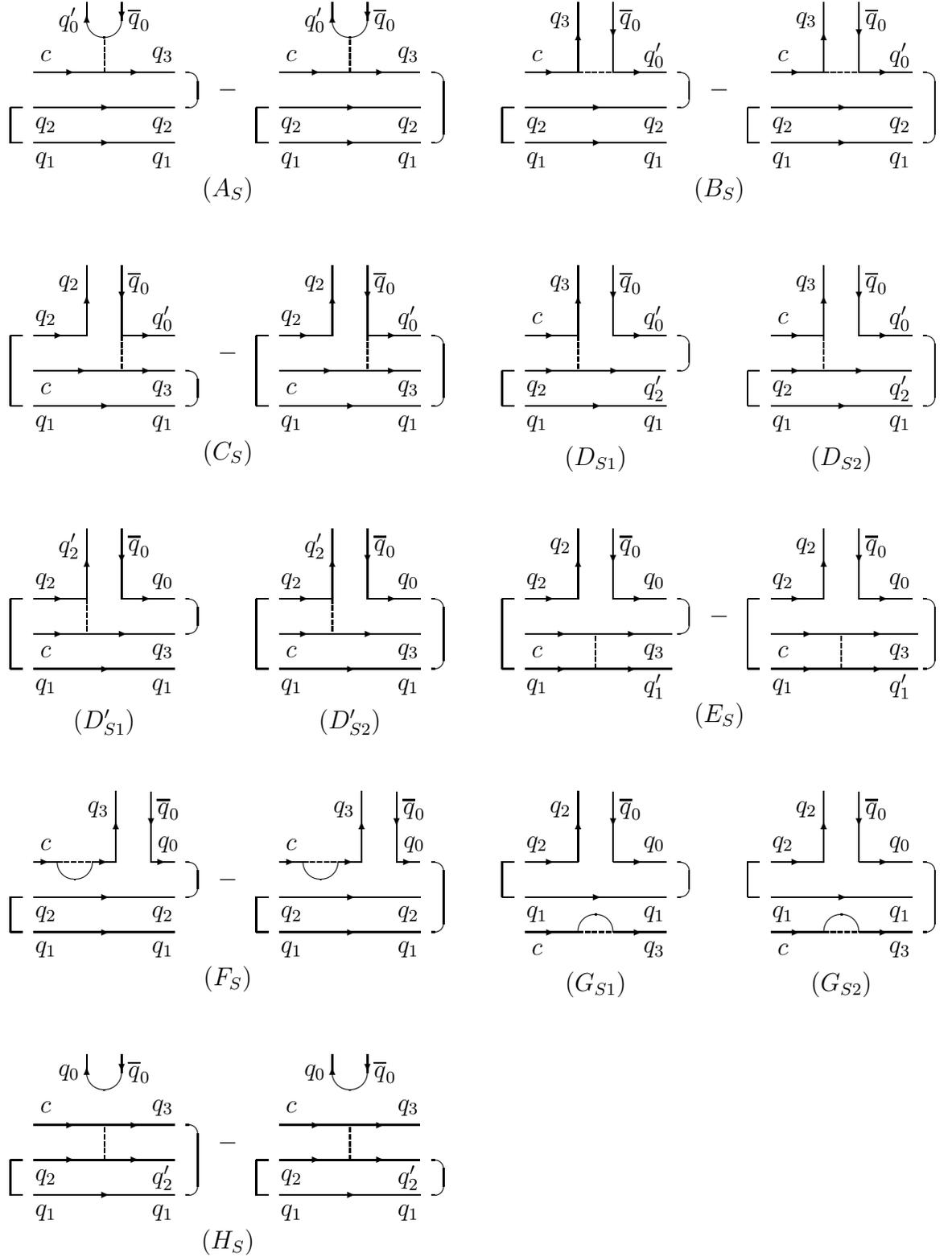
\put(345,10){\makebox(0,0)[b]{Fig. 2. Quark 
diagrams in symmetric bases}}
\end{picture}


\newpage
\begin{thebibliography}{99}
\bibitem{ko} Y. Kohara, Phys. Rev.
   {\bf D44}, 2799 (1991).
\bibitem{ch}L.-L. Chau , H.-Y.Cheng and B. Tseng, Phys. Rev.
   {\bf D54}, 2132 (1996).
\bibitem{k} J.~G.~K\"orner,  Z. Phys. {\bf C43}, 165 (1989).
\bibitem{k2} Y. Kohara, Nihon Univ. preprint (print-95-236) (1995) 
 (unpublished).
\end{thebibliography}
\end{document}